\documentclass[epj]{webofc}
\pdfoutput=1
\usepackage[utf8]{inputenc}
\usepackage[varg]{txfonts}   % Web of Conferences font
\usepackage{booktabs}
\usepackage{xcolor}
\definecolor{darkred}{rgb}{0.4,0.0,0.0}
\definecolor{darkgreen}{rgb}{0.0,0.4,0.0}
\definecolor{darkblue}{rgb}{0.0,0.0,0.4}
\usepackage{subfigure}
\usepackage[justification=justified,
            labelfont=bf,
            font=small,
            singlelinecheck=off]{caption} % enforce justification
\usepackage{graphics}
\wocname{EPJ Web of Conferences}
\woctitle{Lattice2017}
\begin{document}
%%%%%%%%%%%%%%%%%%%%%%%%%%%%%%%%%%%%%%%%%%%%%%%%%%%%%%%%%%%%%%%%%%%%%%%%%%%%%
%
\selectlanguage{english}
%----------------------------------------------------------------------------
\title{%
Non-leptonic kaon decays at large $N_c$\thanks{{\it Presented at the XXXV International Symposium on Lattice Field Theory, Granada (Spain), 18-24 June 2017.}}
}
%----------------------------------------------------------------------------
\author{%
\firstname{Andrea} \lastname{Donini}\inst{1}\fnsep\thanks{Speaker, \email{donini@ific.uv.es}} \and
\firstname{Pilar} \lastname{Hern\'andez}\inst{1} \and
\firstname{Carlos} \lastname{Pena}\inst{2} \and
\firstname{Fernando}  \lastname{Romero-L\'opez}\inst{1,3}
% etc.
}
%----------------------------------------------------------------------------
\institute{%
Instituto de F\'{\i}sica Corpuscular (CSIC-UVEG), E-46071 Valencia, Spain
\and
Departamento  de  F\'{\i}sica  Te\'orica  and  Instituto  de  F\'{\i}sica  Te\'orica  UAM-CSIC,
Universidad  Aut\'onoma  de  Madrid,  E-28049  Madrid,  Spain
\and
Helmholtz Institut f\"ur Strahlen- und Kernphysik, University of Bonn, Bonn, Germany
}
%----------------------------------------------------------------------------
\abstract{%
We study the scaling with the number of colors $N_c$ of the weak amplitudes mediating kaon mixing and decay, in the limit of light charm masses ($m_u=m_d=m_s=m_c$). 
The amplitudes are extracted directly on the lattice for $N_c = 3-7$ (with preliminar results for $N_c = 8$ and $17$) using twisted mass QCD. 
It is shown that the (sub-leading) $1/N_c$ corrections to $\hat B_K$ 
 are small and that the naive $N_c \to \infty$ limit, $\hat B_K = 3/4$, seems to be recovered. On the other hand, the  ${\cal O}\, (1/N_c)$ corrections in $K \to \pi \pi$ amplitudes (derived from $K \to \pi$ matrix elements) are large and fully anti-correlated in the $I=0$ and $I=2$ channels. This may have some implications for the understanding of the $\Delta I=1/2$ rule.
  }
%----------------------------------------------------------------------------
\maketitle
%----------------------------------------------------------------------------
\section{Introduction}\label{intro}

The prediction of flavour violating processes involving kaons remains elusive.  
In particular, there is still no satisfactory explanation of the striking $\Delta I=1/2$ "rule" for which $\Delta I = 1/2$ decays of a kaon into two pions
dominate over $\Delta I = 3/2$ decays (see, however, Refs.~\cite{Boyle:2012ys,Blum:2015ywa}), 
nor a reliable prediction of the parameter that controls direct vs. indirect CP-violation in kaon mixing, $\epsilon'/\epsilon$ \cite{Bai:2015nea}. 
Few attempts have been made at these difficult observables, and the systematic uncertainties in the existing results remain large. At the same time, however, a rather precise determination of the related $K-\bar{K}$ mixing amplitude
(given by $\hat{B}_K$) has emerged \cite{Aoki:2013ldr,Durr:2011ap,Laiho:2011np,Blum:2014tka,Jang:2015sla,Carrasco:2015pra}. 

In \cite{Boyle:2012ys}, the results of the most ambitious lattice computation of $K \rightarrow \pi\pi$ to date were presented, and a significant $\Delta I=1/2$ dominance was observed. It was noted that the $\Delta I=1/2$ rule seems to be originating in an approximate cancellation of the two diagrams (color connected, $C$, and color disconnected, $D$) 
contributing to the $\Delta I=3/2$ amplitude. Roughly speaking, the relative weight of $\Delta I = 1/2$ $K \to \pi \pi$ decays with respect to $\Delta I = 3/2$ ones is 
governed by the ratio $(|C| + |D|)/(|C| - |D|)$. As in Ref.~\cite{Boyle:2012ys} and in its update \cite{Blum:2015ywa} it was found $|C| \sim 0.8 |D|$, a significant enhancement
of this ratio was observed at the non-perturbative level. This result, combined with the known perturbative enhancement of the ratio due to Wilson coefficients 
connecting short- and long-distance QCD matrix elements \cite{Gaillard:1974nj,Altarelli:1974exa}, would then explain the $\Delta I = 1/2$ rule.
Unfortunately, it is not possible to isolate these two contributions physically,  so it is not clear what to extract from this finding. 
In the large $N_c$ expansion \cite{tHooft:1973alw}, however, this is possible since the leading scaling in $N_c$  of the contributions is different.  
The cancellation can therefore be phrased in terms of the sign and size of the $1/N_c$ corrections in the  isospin amplitudes. In fact, it was in the context of phenomenological approaches using the large $N_c$ expansion where the opposite sign of these contributions was first pointed out \cite{Pich:1995qp}. 
Notice that, at the leading order in the $1/N_c$ expansion, the connected diagram $C$ is expected to be $|C| \sim 1/N_c \times |D|$. 
The numerical results of Refs.~\cite{Boyle:2012ys,Blum:2015ywa} are, therefore, pointing to a possible large deviation of the ratio above from the naive $1/N_c$ expectations.
As there is a strong correlation between the $\Delta I=3/2$ amplitude and $\hat{B}_K$, this suggest that the same cancellation 
in the former should be affecting the latter, suggesting a value of $\hat B_K$ significantly smaller  than the $N_c\rightarrow \infty$ value.
A study of the issue of deviations from the na\"{\i}ve factorization approximation
to $K\to\pi\pi$ amplitudes can be found in~\cite{Carrasco:2013jda}.

The large $N_c$ limit of QCD has been  invoked in many phenomenological approaches to this problem  (some relevant references are \cite{Buras:2014maa,Pich:1995qp,Peris:2000sw,Hambye:2003cy,Cirigliano:2011ny}). This seems counter-intuitive since the strict large $N_c$ limit of the $\Delta I=1/2$ rule fails completely. The predictions therefore rely on significant sub-leading $N_c$ effects, a computation that poses formidable difficulties. As a result,  these approaches typically involve 
further approximations beyond  the strict large-$N_c$ expansion. The goal of Ref.~\cite{Donini:2016lwz}, summarized in this proceeding,
was to study from first principles the $N_c$ dependence of  certain $\Delta S=1$ and $\Delta S=2$ amplitudes to check their scaling with the number of colors.

\section{Strategy and simulation details}
\label{sec:strategy}

In order to study the non-perturbative $N_c$ dependence of $K \to \pi \pi$ amplitudes, we have followed the strategy outlined in Ref.~\cite{Giusti:2004an}:
we have measured $K$-$\pi$ and $K$-$\bar{K}$ matrix elements mediated by the four-fermion current-current operators on the lattice, varying the number of colors $N_c$ 
between $N_c = 3$ and $N_c = 7$ (some preliminary results for $N_c = 8$ and $N_c = 17$ will be also shown). In the SU(4)-flavour limit\footnote{We are aware to miss, in this way, the effect of the decoupling of a heavy charm, which was originally argued  to be the origin of the $\Delta I=1/2$ rule \cite{Shifman:1975tn} (something not confirmed by recent non-perturbative studies \cite{Bai:2015nea,Endress:2014ppa}).}, $m_c=m_u=m_d=m_s$, these amplitudes fix $\hat{B}_K$ (up to SU(3) flavour breaking effects by quark masses) and, up to chiral corrections, also the $\Delta I=3/2$ \cite{Donoghue:1982cq,Bijnens:1984ec} and $\Delta I = 1/2$  \cite{Giusti:2004an,Giusti:2006mh} contributions to the non-leptonic kaon decays $K \rightarrow \pi\pi$. The weak Hamiltonian that mediates  CP-conserving $\Delta S=1$ transitions, in terms of four-fermion operators
at the electroweak scale, $\mu \simeq M_W$, takes the following simple form in this limit:
\begin{gather}
\label{eq:heffs1}
H_{\rm w}^{\Delta S=1} = \int d^4x~\frac{g_{\rm w}^2}{4M_W^2}V_{us}^*V_{ud}\sum_{\sigma=\pm} k^\sigma(\mu) \, \bar{Q}^\sigma(x,\mu)\,,
\end{gather}
where $g_{\rm w}^2=4\sqrt{2}G_{\rm F} M_W^2$. Only two four-quark operators of dimension six can appear with the correct symmetry properties under the flavour 
symmetry group  ${\rm SU}(4)_{\rm L} \times {\rm SU}(4)_{\rm R}$, namely
\begin{equation}
\bar Q^\pm(x,\mu) = Z_Q^\pm(\mu) \, \big( J_\mu^{su}(x) J_\mu^{ud}(x) \pm J_\mu^{sd}(x) J_\mu^{uu}(x) \, - \,  [u \leftrightarrow c] \big) \,,
\end{equation}
where $J_\mu$ is the left-handed current, $J_\mu^{\alpha\beta} = (\bar\psi_\alpha\gamma_\mu P_-\psi_\beta)$, 
$P_\pm={1\over 2} (\mathbf{1}\pm\gamma_5)$, and parentheses around quark bilinears indicate that they are traced over spin and colour. 
Eventually, $Z_Q^\pm (\mu)$ is the renormalisation constant of the bare operator $Q^\pm (x)$ computed in some regularisation scheme as, for example, the lattice.
There are other bilinear operators of lower dimensionality that could mix with those above: however, their contribution vanishes in the SU(4) limit  \cite{Giusti:2004an}. 
Notice that this is, indeed,  the limit where the cancellation of  Ref.~\cite{Boyle:2012ys} can be more clearly isolated.   

The operators ${\bar Q}^\sigma(\mu)$ are renormalised at a scale $\mu$ in some renormalisation scheme, being their $\mu$-dependence exactly cancelled by that of the Wilson coefficients $k^\sigma(\mu)$. The renormalisation group invariant (RGI) operators are defined eploiting this fact to eliminate their $\mu$- and scheme-dependence:
\begin{eqnarray}
\hat{Q}^\sigma \equiv \hat{c}^\sigma(\mu) {\bar Q}^\sigma(\mu), \qquad \qquad
{\hat c}^\sigma(\mu)\equiv \left({N_c \over 3} {g^2(\mu)\over 4 \pi} \right)^{\kern-0.2em-{\gamma^\sigma_0\over 2 b_0}}
\kern-1.0em \times \exp\left\{ -\kern-0.3em\int_0^{g(\mu)}\kern-1.4em{\rm d}g
\left[{\gamma^\sigma(g)\over \beta(g)} - {\gamma^\sigma_0\over b_0 \, g}\right]\right\},
\label{eq:c}
\end{eqnarray}
where $g(\mu)$ is the running coupling and $\beta(g)=-g^3 \sum_n b_n g^{2 n}$, $\gamma^\sigma(g) = -g^2 \sum_n \gamma^\sigma_n g^{2 n}$ are the $\beta$-function and the anomalous dimension, respectively. The one- and two-loop coefficients of the $\beta$-function, $b_0$ and $b_1$, and the one-loop coefficient
of the anomalous dimensions, $\gamma_0^\pm$, are renormalisation scheme-independent and can be found in 
Refs.~\cite{Gross:1973id,Politzer:1973fx,Caswell:1974gg,Jones:1974mm,Egorian:1978zx} and \cite{Gaillard:1974nj,Altarelli:1974exa}, respectively.
The normalisation of $\hat{c}^\sigma(\mu)$ coincides with the most popular one for $N_c=3$, whilst using the 't Hooft coupling $\lambda = N_c g^2 (\mu)$
in the first factor instead of the usual coupling, so that the large $N_c$ limit  is well defined.  

We can rewrite the Hamiltonian in terms of RGI quantities, which no longer depend on the scale:
\begin{equation}
\label{eq:WilsonRGI}
\hat{k}^\sigma \equiv {k^\sigma(\mu)\over \hat{c}^\sigma(\mu)}, \qquad \qquad
\hat{k}^\sigma \, \hat{Q}^\sigma  = \left[ { k^\sigma(M_W) \over \hat{c}^\sigma(M_W) } \right] \, \left[ {\hat c}^\sigma(\mu)\, \bar{Q}^\sigma (\mu) \right]
= k^\sigma(M_W) \, U^\sigma(\mu,M_W) \, \bar{Q}^\sigma (\mu) \, ,
\end{equation}
where $\mu$ is a convenient renormalisation scale for the non-perturbative computation of matrix elements of $Q^\pm$, which will be later set to the inverse lattice scale $a^{-1}$. The factor $U^\sigma(\mu, M_W) = {\hat c}^\sigma(\mu)/ {\hat c}^\sigma(M_W)$ measures the running of the renormalised operator between the scales $\mu$ and $M_W$.
In Table~\ref{tab:renorm}  we show the RG running factors needed to compute the renormalised $K \to \pi$ and $K \to \bar K$ matrix elements
as a function of the number of colors. In the evaluation of the $\hat{c}^\sigma(\mu)$ factors we have used the large $N_c$ scaling of $\Lambda_{\rm QCD}$ 
found in Ref.~\cite{Allton:2008ty},
\begin{eqnarray}
{\Lambda_{\overline{MS}}\over \sqrt{\sigma}} = 0.503(2)(40) +{0.33(3)(3)\over N_c^2}.
\label{eq:lambdamsbar}
\end{eqnarray}
The values of the normalisation coefficients $\hat c^\pm (a^{-1})$ and of the running of the renormalised operators from the scale of lattice computations, $\mu = a^{-1}$, 
to the scale of the effective theory, $M_W$, have been computed using perturbative results at two loops in the RI scheme \cite{Ciuchini:1997bw,Buras:2000if}.
This implies relying on perturbation theory at scales above $\mu=a^{-1} \sim 2~{\rm GeV}$.

\begin{table}[thb]
  \small
  \centering
  \caption{Perturbative RG running factors.  $U^\sigma$ and $k^\sigma$ are computed using the two-loop $\overline{\rm MS}$ coupling (with $\Lambda_{\overline{MS}}$ 
  taken from eq.~(\ref{eq:lambdamsbar}) from ref.~\cite{Allton:2008ty}).
  }
  \label{tab:renorm}
  \begin{tabular}{c@{\hspace{5mm}}cccccccc}
\hline\hline\\[-2.0ex]
$N_c$ &
$\hat{k}^+$ & $k^+(M_W)$ & $U^+(a^{-1},M_W)$ & $\hat{c}^+(a^{-1})$ &  $\hat{k}^-$ & $k^-(M_W)$ & $U^-(a^{-1},M_W)$ & $\hat{c}^-(a^{-1})$ \\[0.3ex] 
\hline\\[-2.0ex]
 3 & 0.642 & 1.030 & 0.875 & 1.404  & 2.398 & 0.940 & 1.319 & 0.517 \\
 4 & 0.658 & 1.025 & 0.895 & 1.394  & 1.998 & 0.958 & 1.210 & 0.580 \\
 5 & 0.679 & 1.021 & 0.910 & 1.368  & 1.780 & 0.968 & 1.156 & 0.620 \\
 6 & 0.700 & 1.018 & 0.921 & 1.340  & 1.643 & 0.974 & 1.124 & 0.666 \\
 7 & 0.719 & 1.016 & 0.930 & 1.315  & 1.550 & 0.978 & 1.103 & 0.696 \\
 \hline
 8   & 0.736 & 1.015  & 0.938  & 1.293  & 1.480 & 0.981 & 1.088  & 0.721  \\
 17 & 0.827 & 1.007 & 0.968   & 1.178 & 1.238 & 0.992   & 1.038   & 0.832 \\
\hline\hline
\end{tabular}
 \end{table}

As specified above, our goal is to compute the $K\to\pi$ amplitudes mediated by $H_{\rm w}^{\Delta S=1}$.
The hadronic contribution is encoded in the ratios of the following matrix elements:
\begin{eqnarray}
\hat{R}^\pm \equiv \frac{\langle\pi|\hat{Q}^\pm|K\rangle}{f_K f_\pi m_Km_\pi}
= \hat{c}^\pm(\mu) Z_R^\pm(\mu) R^\pm \,,
\label{eq:ratio}
\end{eqnarray}
where $Z_R^\pm(\mu)$ are the renormalisation factors for the ratios and $R^\pm$ is the ratio of matrix elements of bare operators. 
The ratio of the two isospin amplitudes $i A_{0,2} e^{i \delta_{0,2}} \equiv \langle (\pi\pi)_{0,2} | H_W |K_0\rangle$ (where the subindex refers to the final isospin state)
can be related in chiral perturbation theory in the GIM limit to the $K\to\pi$
amplitudes $A^\pm  \equiv \hat{k}^\pm \hat{R}^\pm$  \cite{Giusti:2004an} as follows: 
\begin{eqnarray}
{A_0 \over A_2} = {1 \over \sqrt{2}} \left({1\over 2} + {3 \over 2} {A^-\over A^+}\right) \, ,
\end{eqnarray} 
from which we can see that the large enhancement of the ratio $|A_0/A_2| \sim 22$ is related, in this limit, to a large value of the ratio of the amplitudes $A^-/A^+$
(up to chiral corrections).  In the SU(3) limit $m_s=m_d=m_u$, the kaon mixing amplitude given by $\hat B_K$ is also related to $R^+$, $\hat{B}_K = {3 \over 4} \hat{R}^+ $
(something not true outside of the SU(3) limit, due to large chiral corrections \cite{Donoghue:1982cq,Bijnens:1984ec}).
The relation between $K \to \pi \pi$ and $K \to \pi$ amplitudes, computed up to one loop in ChPT in the leading-log approximation, is:
\begin{eqnarray}
\left.{\langle \pi^+ \pi^0| H_W | K\rangle \over m_K^2 -m_\pi^2}\right|_{ m_s = m_d} = {i F\over \sqrt{2}} A^+ G_F V_{ud} V_{us}^*, 
\end{eqnarray}
where $F$ is the decay constant in the chiral limit and $A^+$ contains one loop corrections. This shows that, in this approximation,  the $1/N_c$ corrections 
in the physical amplitude are fixed\footnote{It has been argued, however, that higher-order ChPT effects may have an important impact on
$K\to\pi\pi$ amplitudes at the same order in $1/N_c$ (see, {\em e.g.}, Refs.~\cite{Truong:1987hn,Isgur:1989js,Kambor:1991ah,Pallante:2000hk}).} by those in $A^+$. 

On the lattice, the ratios $\hat R^\pm$ are extracted as follows:
\begin{eqnarray}
\label{eq:bareratios}
R^\pm  = \kern-1.0em
\lim_{ \substack{z_0-x_0\to\infty \\ y_0-z_0\to \infty}}
\frac{\sum_{{\mathbf x},{\mathbf y}}\langle P^{du}(y) Q^\pm(z) P^{us}(x)\rangle}
{\sum_{{\mathbf x,\mathbf y}}\langle P^{du}(y)A_0^{ud}(z)\rangle \langle A_0^{su}(z) P^{us}(x)\rangle}\,,
\end{eqnarray}
where $P^{ab}(x)=\bar{\psi}^a(x) \gamma_5\psi^b(x)$, and $A^{ab}_0(x)=Z_{\rm A}\bar{\psi}^a(x) \gamma_0 \gamma_5\psi^b(x)$. 
We have computed the renormalised ratios $\hat R^\pm$ in the quenched approximation. This does not modify the leading large $N_c$ result, but it can modify 
the first sub-leading $1/N_c$ corrections (we plan to address this issue in further studies). 
We have implemented the required correlation functions in the source code first developed in \cite{DelDebbio:2008zf} and further optimized in \cite{pica}. 
The number of colors and the lattice size in the time direction are given in the first two columns of Table~\ref{tab:sim}. The spatial volume, $L/a = 16$, is kept fixed in all simulations
(but for $N_c = 17$, for which $L/a = 12$, as in Ref.~\cite{Bali:2013kia}). Following \cite{Bali:2013kia} the bare coupling, $\beta= 2 N_c/g_0^2$,  is tuned with $N_c$ in such a way that the string tension is $a \sqrt{\sigma} \simeq 0.2093$ for all $N_c$; this results in $a \simeq 0.093~{\rm fm}$ with $\sigma=1~{\rm GeV/fm}$. 
The bare 't Hooft coupling $\lambda$ for $N_c \in [3,7]$ is well described by the scaling:
\begin{eqnarray}
\lambda = N_c g^2_0 = 2.775(3) +{1.90(3)\over N_c^2}.
\end{eqnarray}
The coupling $\beta$ as a function of $N_c$ is given in the third column of Table~\ref{tab:sim}. The gauge action is the standard plaquette action. On the other hand, 
in order to preserve the multiplicative renormalisation of $Q^\pm$, while avoiding the high computational cost of a simulation with exactly chiral lattice fermions, 
we use a Wilson twisted-mass fermion regularisation \cite{Frezzotti:2000nk,Frezzotti:2003ni}. This allows to devise a formulation of valence quarks that not only
preserves good renormalisation properties, but also prevents the appearance of linear cutoff effects in $a$ \cite{Frezzotti:2004wz}.
The full-twist condition amounts to having a vanishing current quark mass $m_{\rm PCAC}$ from the axial
Takahashi-Ward identity in so-called twisted quark field variables. The value of $a m_{\rm PCAC}$ in our simulations is given in the fourth column of Table~\ref{tab:sim}, 
where we can see that the full-twist condition $a m_{\rm PCAC} = 0$, expected from an accurate tuning of the Wilson
critical mass (which we again take from \cite{Bali:2013kia}), is satisfied to a varying degree of
accuracy; the deviations present are however irrelevant within the precision of our results.
The bare quark mass is chosen to provide a pseudoscalar mass not far from the physical kaon mass in all cases (see the fifth column of Table~\ref{tab:sim}).

\begin{table}[h]
  \small
  \centering
  \caption{Lattice simulation results. Lattice sizes are $(L/a)^3 \times (T/a)$, with $L/a=16$ throughout ($^\star$ with the only exception of $N_c = 17$, for which $L/a = 12$). 
  The twisted bare mass is fixed to $a \mu=0.02$. The lattice spacing is fixed by the string tension: $a \sqrt{\sigma} \simeq 0.2093$ \cite{Bali:2013kia}. $m_{\rm PCAC}$ 
is the current mass obtained from the axial Takahashi-Ward identity in twisted quark field variables. $m_{\rm PS}$
is the meson mass in the SU(3) limit. $R^\pm$ are our results for the bare ratios given in eq.~(\ref{eq:bareratios}).
$Z^\sigma (a^{-1})$ at one-loop have been extracted from \cite{Constantinou:2010zs,Alexandrou:2012mt}.}
  \label{tab:sim}
  \begin{tabular}{clcr@{\hspace{0mm}}c@{\hspace{0mm}}lcllcc}
\hline\hline
$N_c$ & 
$T/a$ &
~~~~$\beta$ &
\multicolumn{3}{c}{$a m_{\rm\scriptscriptstyle PCAC}$} &  $a m_{\rm PS} 
$ & ~~$R^+_{\rm bare}$ & ~~$R^-_{\rm bare}$  & $Z^+(a^{-1})$ & $Z^-(a^{-1})$ \\[0.5ex] 
\hline\\[-2.0ex]
3 & $48$ & 6.0175  & -0&.&002(14)  & 0.2718(61) & 0.774(21) & 1.218(31) & 0.983 &  1.059 \\
4 & $48$ & 11.028  & -0&.&0015(11) & 0.2637(39) & 0.783(15) & 1.198(19)  & 0.988 & 1.043 \\
5 & $48$ & 17.535  &  0&.&0028(9)  & 0.2655(31) & 0.839(8)  & 1.145(12) & 0.991 & 1.035 \\
6 & $32$ & 25.452  &  0&.&0013(7)  & 0.2676(28) & 0.871(6)  & 1.125(7)  & 0.994 & 1.030 \\
7 & $32$ & 34.8343 & -0&.&0034(6)  & 0.2819(19) & 0.880(5)  & 1.122(5) & 0.996 & 1.026 \\
\hline
8 & $32$ & 45.7003 &  -0&.&0002(6) & 0.3065(23) & 0.872(4) &  1.127(4) & 0.997 & 1.024 \\
17 & $24^\star$ & 208.45 &  -0&.&00178(45) & 0.2594(9) & 0.967(3) & 1.039(6) & 1.003 & 1.015 \\
 \hline
\hline\end{tabular}
\end{table}

Eventually, our results for the bare ratios $R^\pm$ defined in eq.~(\ref{eq:bareratios}), computed  in the SU(3) limit, are shown in the sixth and seventh columns of the table. 
The corresponding renormalisation constants, $Z^\pm$, have been computed at one-loop in the RI scheme using the scripts provided in Refs.~\cite{Constantinou:2010zs,Alexandrou:2012mt}.
Notice that, due to the breaking of chiral symmetry in the adopted regularisation, the axial current requires
a finite, $N_c$-dependent, renormalisation constant $Z_A$ (taken from the same references), that has to be included in the factors $Z^\pm$.   
The values of $Z^\pm (a^{-1})$ are given in the two rightmost columns of Table~\ref{tab:sim}.
The results shown for $R^\pm$ correspond to ${\cal O}(100)$ measurements at each value of $N_c$, with each measurement taken every 1000 gauge updates. 
The only exception is the run for $N_c = 17$, for which we have only 17 measurements taken every 100 gauge updates. We have checked, however, that all measurements
are fully decorrelated using the techniques of Ref.~\cite{Wolff:2003sm}.

\section{Results}
\label{sec:results}

 \begin{figure}[thb]
  \centering
  \includegraphics[width=7cm,clip]{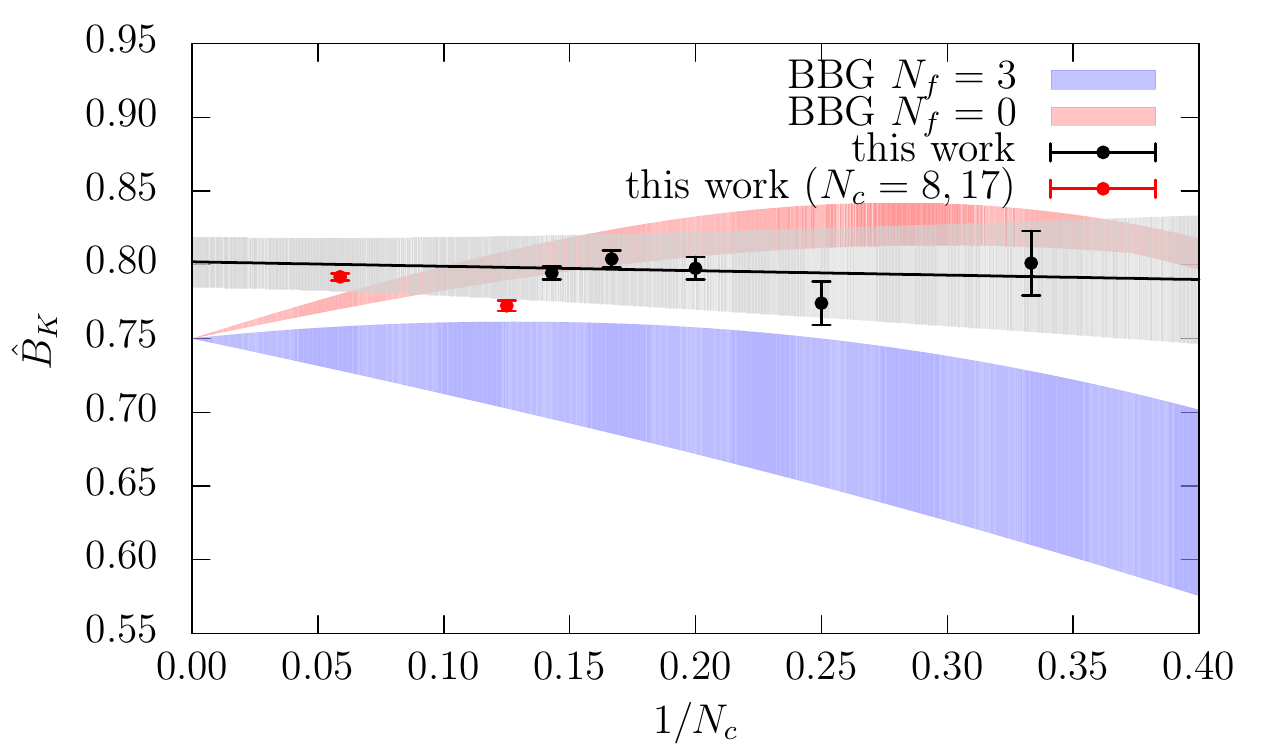}
   \includegraphics[width=7cm,clip]{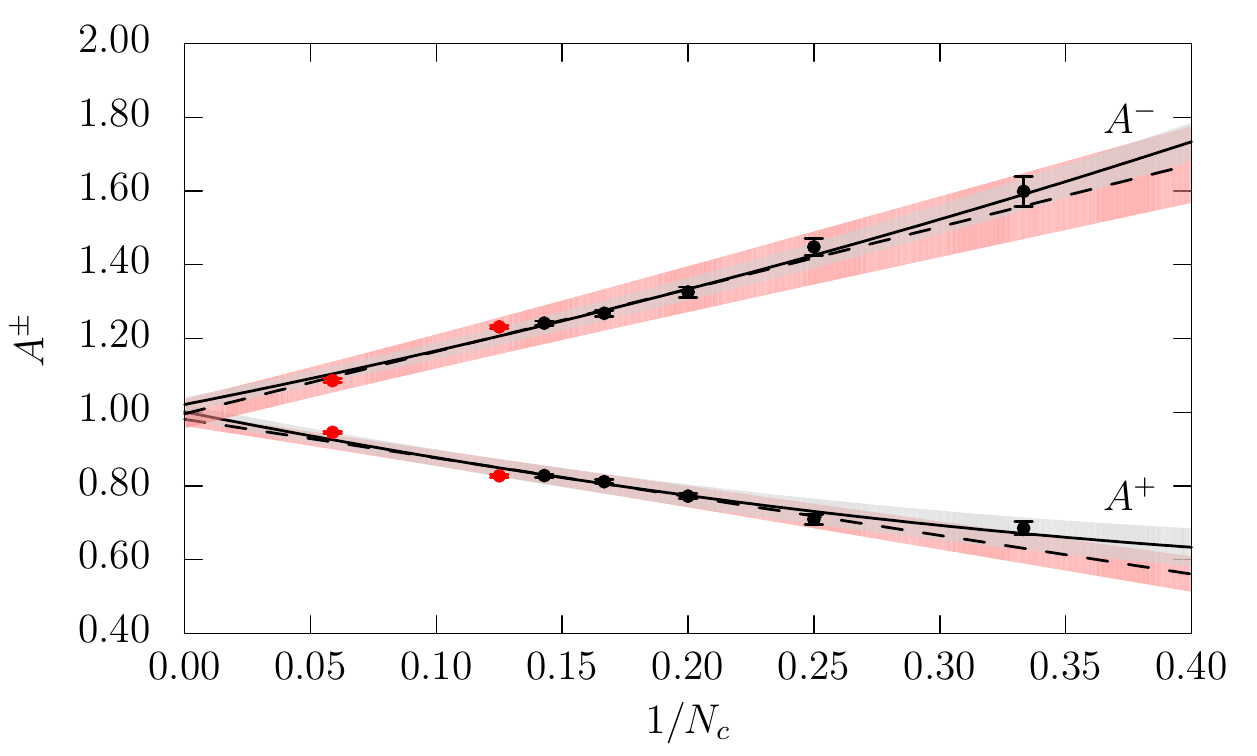}
  \caption{Left: 
  $\hat{B}_K$ versus $1/N_c$.
The grey band (solid line) is a linear fit to our five data points. The red and blue bands use the model prediction of~\cite{Buras:2014maa}.
Right: $A^\pm$ versus $1/N_c$. 
The grey bands (solid lines) are obtained from the results of the fits to $1/2 (A^- \pm A^+)$ in eqs.~(\ref{eq:fit}); the red bands (dashed lines) are linear fits including $N_c=4-7$ from Table~\ref{tab:fits}. Data points for $N_c = 8$ and 17 (in red) have been added to the plots, but not used in the fits.
}
  \label{fig:bkandApm}
\end{figure}

Our results for $\hat{B}_K$ as a function of $1/N_c$ are shown in Fig.~\ref{fig:bkandApm}(left) together with a linear fit to the data
for $N_c = 3-7]$, represented by a solid black line (results for $N_c = 8$ and 17 have not been used in the fit). The grey band 
shows the 1$\sigma$ error on the fit. The parameters of the fit are shown in the first two lines of Table~\ref{tab:fits} for two choices of the data points included, 
together with the corresponding $p$-values. The third line of the same table shows our result for a quadratic fit to the data. 
We compare our results with our own evaluation of the predictions of the phenomenological analysis in Ref.~\cite{Buras:2014maa}, 
represented by a light red band for $N_f = 3$ and by a blue band for $N_f = 0$. For $N_f=3$ we use the same values for hadronic masses and decay constants as 
in~\cite{Buras:2014maa}, and derive the decay constant for $N_c \neq 3$ by rescaling $F_K=F_K(N_c=3)\sqrt{N_c/3}$. For $N_f=0$ we use as input for the hadronic quantities, including their $N_c$ dependence, the interpolating formulae provided in~\cite{Bali:2013kia}, matched to our measured values of $M_K$. The band represent the difference 
between setting the matching scale $M$ in eq.~(62) of~\cite{Buras:2014maa} at $0.6~{\rm GeV}$ and at $1~{\rm GeV}$; for $N_f=0$ it also comprises the uncertainty due to our value of $M_K$ not being constant within errors as a function of $N_c$.
Notice that both theoretical predictions give $\hat B_K = 3/4$ in the $N_c \to \infty$ limit. On the other hand, our data (and the preliminar values for $N_c = 8$ and 17) give
a value for $\hat B_K$ at $N_c \to \infty$ slightly larger than the theoretical expectation. Errors are, however, still too large to draw any conclusion. For example, 
a significant $O(a^2)$ uncertainty for $R^+$ can be expected, cf. the $\mathcal{O}(10\%)$ effect for $N_c=3$, $N_f=2$ shown by the data of \cite{Constantinou:2010qv} 
at a lattice spacing comparable to ours.

\begin{table}[thb]
  \small
     \centering
  \caption{
  Fit parameters of $A^\pm$ assuming a linear (l) or quadratic (q) dependence, and various fit ranges (using only data up to $N_c = 7$, though).
The order at which each coefficient enters in the polynomial ansatz in powers of $1/N_c$ is indicated, alongside with the $p$-value for each fit.
 }
  \label{tab:fits}
  \begin{tabular}{c@{\hspace{5mm}}c@{\hspace{5mm}}ccc@{\hspace{5mm}}c}
\hline
\hline
obs & fit & $1$ & $1/N_c$ & $1/N_c^2$ & $p$-value \\[0.3ex] 
\hline\\[-2.5ex]
$\hat{B}_K$  & l, $N_c \geq 3$ & 0.802(17) & -0.03(10) & --- & 0.24 \\
             & l, $N_c \geq 4$ & 0.808(27) & -0.07(16) & --- & 0.14 \\
             & q, $N_c \geq 3$ & 0.788(79) &  0.12(78) & -0.3(1.8) & 0.12 \\[0.3ex]
\hline\\[-2.5ex]
$A^+$        & l, $N_c \geq 3$ & 0.956(20) & -0.89(11) & --- & 0.10 \\
             & l, $N_c \geq 4$ & 0.981(18) & -1.05(11) & --- & 0.39 \\[0.3ex]
\hline\\[-2.5ex]
$A^-$        & l, $N_c \geq 3$ & 0.984(28) &  1.77(17) & --- & 0.21 \\
             & l, $N_c \geq 4$ & 0.996(39) &  1.69(24) & --- & 0.14 \\[0.3ex]
 \hline
\hline
\end{tabular}
 \end{table}
 
From Fig.~\ref{fig:bkandApm}(left) we can see that the sub-leading $1/N_c$ corrections in $\hat{B}_K$ are small (which goes in the direction of the predictions in \cite{Buras:2014maa}, but not those in \cite{Peris:2000sw}, that correspond to the chiral limit). The smallness of $1/N_c$ corrections in $\hat{B}_K$ is related to the RGI normalization of this quantity, $\hat c^+(a^{-1})$: the $N_c$-dependence of $R^+$ (see Table~\ref{tab:sim}) is cancelled by the RGI Wilson coefficient $\hat k^+$ (see Table~\ref{tab:renorm}).
In contrast,  the total $K\to\pi$ amplitudes  show  very significant sub-leading $1/N_c$ corrections, as shown in Fig.~\ref{fig:bkandApm}(right).
In the Figure we present our data for $A^\pm$ and the results of a linear (dashed lines) and quadratic (solid lines)
fit to the data, obtained using again only data for $N_c = 3-7$ (preliminar data for $N_c = 8$ and 17, although not used in the fits, are perfectly compatible with the 
results obtained for lower values of $N_c$). The parameters of the linear fit for $A^+$ and $A^-$ are shown in the fourth and fifth (sixth and seventh) lines of Table~\ref{tab:fits}, respectively. We can see that $A^+$ and $A^`$ are strongly anti-correlated in $N_c$ and that their extrapolation at $N_c \to \infty$ is in very good 
agreement with theoretical expectations (for which $|A_0/A_2|_{N_c \to \infty} \sim \sqrt{2}$). Notice that, in the GIM limit, the chiral logs have been shown to be fully anti-correlated in $A^\pm$ \cite{Hernandez:2006kz} and therefore an extrapolation to the chiral limit using chiral perturbation theory should not change the anti-correlation found here. Unfortunately, the computation of chiral logs  in $K\rightarrow (\pi\pi)_{I=0}$ in the GIM limit is not yet available.

\begin{figure}[thb]
  \centering
    \includegraphics[width=7cm,clip]{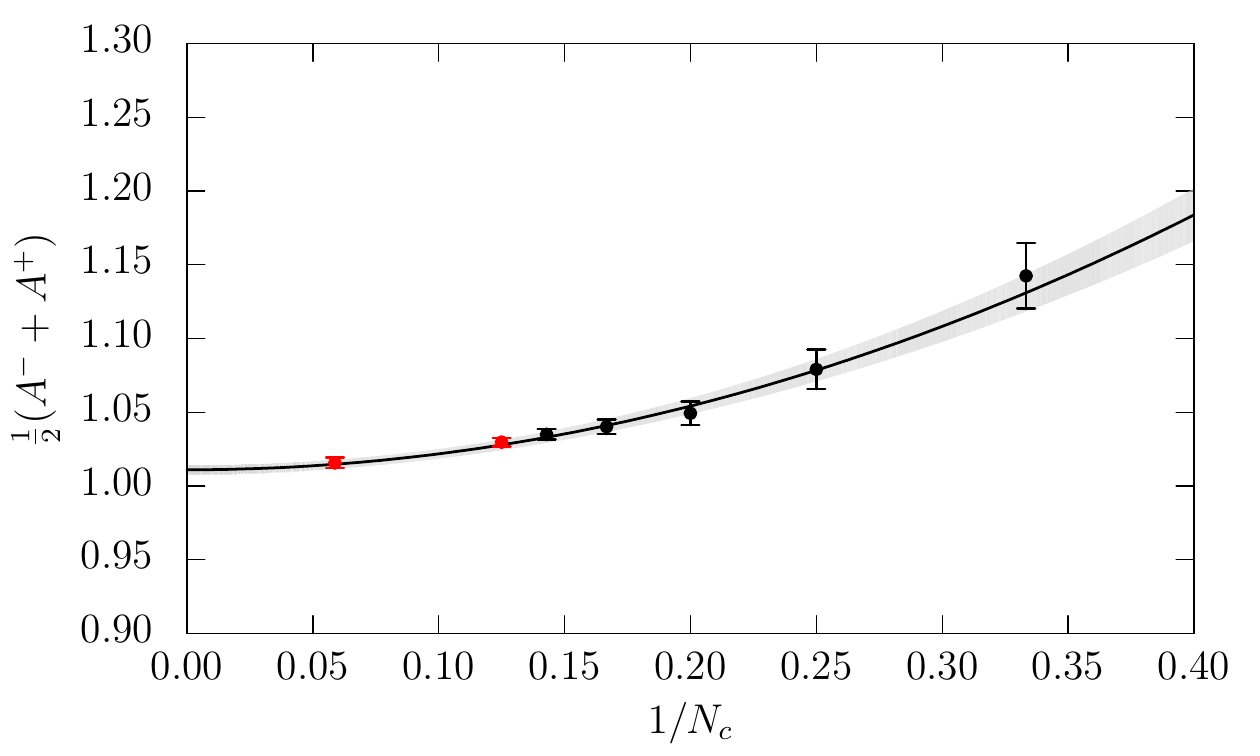} 
  \includegraphics[width=7cm,clip]{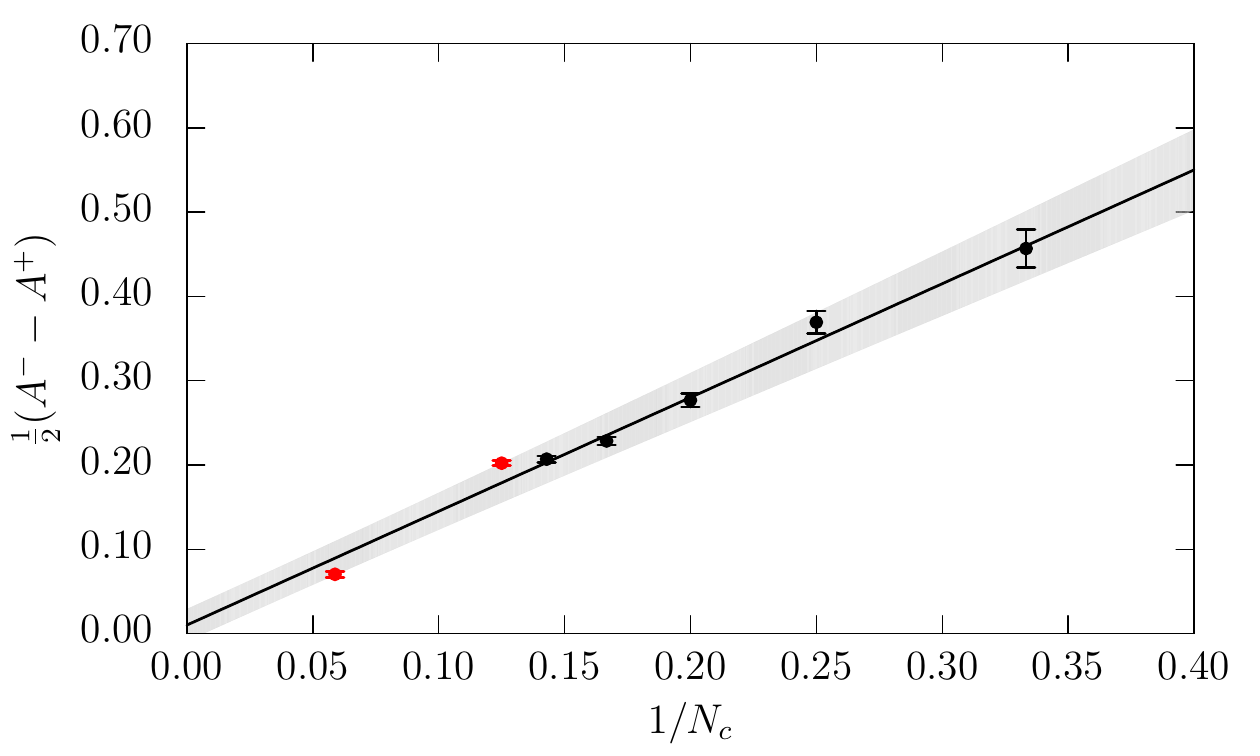}
    \caption{
  ${A^-\pm A^+\over 2}$ versus $1/N_c$. The bands (solid lines) are quadratic and linear fits in $1/N_c$, respectively.
  Data points for $N_c = 8$ and 17 (in red) have been added to the plots, but not used in the fits.
 }
  \label{fig:comb}
\end{figure}

Eventually, our results for the combinations ${1\over 2}(A^-\pm A^+)$, corresponding to the (renormalized) connected $\hat C$ and disconnected $\hat D$ diagrams, respectively, are shown in Fig.~\ref{fig:comb}. A quadratic fit using $N_c = 3-7$ data gives the following results: \begin{gather}
\begin{split}
\hat C = {A^--A^+\over 2} &= 0.01(2) + {1.35(11) \over N_c}   \quad (p{\rm -value} =0.12), \\ 
\hat D = {A^-+A^+\over 2} &= 1.01(3) + {1.08(11) \over N^2_c} \quad (p{\rm -value} =0.81).
\end{split}
\label{eq:fit}
\end{gather}
Our results show that the sub-leading $1/N_c$ effects cancel in the "disconnected" contribution to $K \to \pi$, whereas they are the only visible corrections in the 
"connected" one.  In particular, it is clear that a relation as the one expected naively in the large $N_c$ expansion, $|\hat C| \sim k/N_c \times |\hat D|$, holds, 
with a coefficient $k \sim 1.3$ much smaller than what found in Refs.~\cite{Boyle:2012ys,Blum:2015ywa}. The source of the huge non-perturbative cancellation
between "connected" and "disconnected" contributions seem not to arise from a failure of naive $N_c$ scaling, but in some other enhancement of the coefficient $k$
that relates the two amplitudes (for example, large $1/N_c$ corrections could be present at the physical point, $m_s \gg m_d$, as suggested by a large chiral log).

We have not included any systematic error in these results. There are two obvious sources: finite lattice spacing and the quenched approximation. Although 
it is impossible to quantify those errors, we do not expect them to be larger that those observed at $N_c=3$, where they have been studied.
The pioneering large-$N_c$ study of dynamical QCD in~\cite{DeGrand:2016pur} shows that an extension
of our work to take into account unquenching effects is feasible.

\section*{Acknowledgments}
We warmly thank  C.~Pica for providing us with a $SU(N_c)$ lattice code. 
This work was partially supported by grants FPA2012-31686, FPA2014-57816-P, FPA2015-68541-P (MINECO/FEDER), PROMETEOII/2014/050, MINECO's ``Centro de Excelencia Severo Ochoa'' Programme under grants SEV-2012-0249 and SEV-2014-0398, 
and the European projects H2020-MSCA-ITN-2015//674896-ELUSIVES and H2020-MSCA-RISE-2015. 

%\clearpage
\bibliography{Lattice2017}

\begin{thebibliography}{49}

\bibitem{Boyle:2012ys}
P.A. Boyle et~al. (RBC, UKQCD), Phys. Rev. Lett. \textbf{110}, 152001 (2013),
  \texttt{1212.1474}

\bibitem{Blum:2015ywa}
T.~Blum et~al., Phys. Rev. \textbf{D91}, 074502 (2015), \texttt{1502.00263}

\bibitem{Bai:2015nea}
Z.~Bai et~al. (RBC, UKQCD), Phys. Rev. Lett. \textbf{115}, 212001 (2015),
  \texttt{1505.07863}

\bibitem{Aoki:2013ldr}
S.~Aoki et~al., Eur. Phys. J. \textbf{C74}, 2890 (2014), \texttt{1310.8555}

\bibitem{Durr:2011ap}
S.~Durr et~al., Phys. Lett. \textbf{B705}, 477 (2011), \texttt{1106.3230}

\bibitem{Laiho:2011np}
J.~Laiho, R.S. Van~de Water, PoS \textbf{LATTICE2011}, 293 (2011),
  \texttt{1112.4861}

\bibitem{Blum:2014tka}
T.~Blum et~al. (RBC, UKQCD), Phys. Rev. \textbf{D93}, 074505 (2016),
  \texttt{1411.7017}

\bibitem{Jang:2015sla}
B.J. Choi et~al. (SWME), Phys. Rev. \textbf{D93}, 014511 (2016),
  \texttt{1509.00592}

\bibitem{Carrasco:2015pra}
N.~Carrasco et~al. (ETM), Phys. Rev. \textbf{D92}, 034516 (2015),
  \texttt{1505.06639}

\bibitem{Gaillard:1974nj}
M.K. Gaillard, B.W. Lee, Phys. Rev. Lett. \textbf{33}, 108 (1974)

\bibitem{Altarelli:1974exa}
G.~Altarelli, L.~Maiani, Phys. Lett. \textbf{52B}, 351 (1974)

\bibitem{tHooft:1973alw}
G.~'t~Hooft, Nucl. Phys. \textbf{B72}, 461 (1974)

\bibitem{Pich:1995qp}
A.~Pich, E.~de~Rafael, Phys. Lett. \textbf{B374}, 186 (1996),
  \texttt{hep-ph/9511465}

\bibitem{Carrasco:2013jda}
N.~Carrasco, V.~Lubicz, L.~Silvestrini (ETM), Phys. Lett. \textbf{B736}, 174
  (2014), \texttt{1312.6691}

\bibitem{Buras:2014maa}
A.J. Buras, J.M. Gerard, W.A. Bardeen, Eur. Phys. J. \textbf{C74}, 2871 (2014),
  \texttt{1401.1385}

\bibitem{Peris:2000sw}
S.~Peris, E.~de~Rafael, Phys. Lett. \textbf{B490}, 213 (2000),
  \texttt{hep-ph/0006146}

\bibitem{Hambye:2003cy}
T.~Hambye, S.~Peris, E.~de~Rafael, JHEP \textbf{05}, 027 (2003),
  \texttt{hep-ph/0305104}

\bibitem{Cirigliano:2011ny}
V.~Cirigliano et~al., Rev. Mod. Phys. \textbf{84}, 399 (2012),
  \texttt{1107.6001}

\bibitem{Donini:2016lwz}
A.~Donini, P.~Hernandez, C.~Pena, F.~Romero-Lopez, Phys. Rev. \textbf{D94},
  114511 (2016), \texttt{1607.03262}

\bibitem{Giusti:2004an}
L.~Giusti et~al., JHEP \textbf{11}, 016 (2004), \texttt{hep-lat/0407007}

\bibitem{Shifman:1975tn}
M.A. Shifman, A.I. Vainshtein, V.I. Zakharov, Nucl. Phys. \textbf{B120}, 316
  (1977)

\bibitem{Endress:2014ppa}
E.~Endress, C.~Pena, Phys. Rev. \textbf{D90}, 094504 (2014), \texttt{1402.0827}

\bibitem{Donoghue:1982cq}
J.F. Donoghue, E.~Golowich, B.R. Holstein, Phys. Lett. \textbf{119B}, 412
  (1982)

\bibitem{Bijnens:1984ec}
J.~Bijnens, H.~Sonoda, M.B. Wise, Phys. Rev. Lett. \textbf{53}, 2367 (1984)

\bibitem{Giusti:2006mh}
L.~Giusti et~al., Phys. Rev. Lett. \textbf{98}, 082003 (2007),
  \texttt{hep-ph/0607220}

\bibitem{Gross:1973id}
D.J. Gross, F.~Wilczek, Phys. Rev. Lett. \textbf{30}, 1343 (1973)

\bibitem{Politzer:1973fx}
H.D. Politzer, Phys. Rev. Lett. \textbf{30}, 1346 (1973)

\bibitem{Caswell:1974gg}
W.E. Caswell, Phys. Rev. Lett. \textbf{33}, 244 (1974)

\bibitem{Jones:1974mm}
D.R.T. Jones, Nucl. Phys. \textbf{B75}, 531 (1974)

\bibitem{Egorian:1978zx}
E.~Egorian, O.V. Tarasov, Teor. Mat. Fiz. \textbf{41}, 26 (1979), [Theor. Math.
  Phys.41,863(1979)]

\bibitem{Allton:2008ty}
C.~Allton, M.~Teper, A.~Trivini, JHEP \textbf{07}, 021 (2008),
  \texttt{0803.1092}

\bibitem{Ciuchini:1997bw}
M.~Ciuchini et~al., Nucl. Phys. \textbf{B523}, 501 (1998),
  \texttt{hep-ph/9711402}

\bibitem{Buras:2000if}
A.J. Buras, M.~Misiak, J.~Urban, Nucl. Phys. \textbf{B586}, 397 (2000),
  \texttt{hep-ph/0005183}

\bibitem{Truong:1987hn}
T.N. Truong, Phys. Lett. \textbf{B207}, 495 (1988)

\bibitem{Isgur:1989js}
N.~Isgur, K.~Maltman, J.D. Weinstein, T.~Barnes, Phys. Rev. Lett. \textbf{64},
  161 (1990)

\bibitem{Kambor:1991ah}
J.~Kambor, J.H. Missimer, D.~Wyler, Phys. Lett. \textbf{B261}, 496 (1991)

\bibitem{Pallante:2000hk}
E.~Pallante, A.~Pich, Nucl. Phys. \textbf{B592}, 294 (2001),
  \texttt{hep-ph/0007208}

\bibitem{DelDebbio:2008zf}
L.~Del~Debbio, A.~Patella, C.~Pica, Phys. Rev. \textbf{D81}, 094503 (2010),
  \texttt{0805.2058}

\bibitem{pica}
C.~Pica (2016), {private communication.}

\bibitem{Bali:2013kia}
G.S. Bali et~al., JHEP \textbf{06}, 071 (2013), \texttt{1304.4437}

\bibitem{Frezzotti:2000nk}
R.~Frezzotti, P.A. Grassi, S.~Sint, P.~Weisz (Alpha), JHEP \textbf{08}, 058
  (2001), \texttt{hep-lat/0101001}

\bibitem{Frezzotti:2003ni}
R.~Frezzotti, G.C. Rossi, JHEP \textbf{08}, 007 (2004),
  \texttt{hep-lat/0306014}

\bibitem{Frezzotti:2004wz}
R.~Frezzotti, G.C. Rossi, JHEP \textbf{10}, 070 (2004),
  \texttt{hep-lat/0407002}

\bibitem{Constantinou:2010zs}
M.~Constantinou et~al., Phys. Rev. \textbf{D83}, 074503 (2011),
  \texttt{1011.6059}

\bibitem{Alexandrou:2012mt}
C.~Alexandrou et~al., Phys. Rev. \textbf{D86}, 014505 (2012),
  \texttt{1201.5025}

\bibitem{Wolff:2003sm}
U.~Wolff (ALPHA), Comput. Phys. Commun. \textbf{156}, 143 (2004), [Erratum:
  Comput. Phys. Commun.176,383(2007)], \texttt{hep-lat/0306017}

\bibitem{Constantinou:2010qv}
M.~Constantinou et~al. (ETM), Phys. Rev. \textbf{D83}, 014505 (2011),
  \texttt{1009.5606}

\bibitem{Hernandez:2006kz}
P.~Hernandez, M.~Laine, JHEP \textbf{10}, 069 (2006), \texttt{hep-lat/0607027}

\bibitem{DeGrand:2016pur}
T.~DeGrand, Y.~Liu, Phys. Rev. \textbf{D94}, 034506 (2016), [Erratum: Phys.
  Rev.D95,no.1,019902(2017)], \texttt{1606.01277}

\end{thebibliography}

%%%%%%%%%%%%%%%%%%%%%%%%%%%%%%%%%%%%%%%%%%%%%%%%%%%%%%%%%%%%%%%%%%%%%%%%%%%%%
\end{document}